\documentclass[sn-mathphys-num,iicol]{sn-jnl}% Math and Physical Sciences Numbered Reference Style 
\usepackage{graphicx}%
\usepackage{epstopdf}
\usepackage{multirow}%
\usepackage{amsmath,amssymb,amsfonts}%
\usepackage{amsthm}%
\usepackage{mathrsfs}%
\usepackage{xcolor}%
\usepackage{textcomp}%
\usepackage{manyfoot}%
\usepackage{booktabs}%
\usepackage{algorithm}%
\usepackage{algorithmicx}%
\usepackage{algpseudocode}%
\usepackage{listings}%
\usepackage{siunitx}
\usepackage{subcaption}
\usepackage{float}
\usepackage{stfloats}
\begin{document}

\title{Parallelized Event Data Management System Based on MT-SNiPER Framework and PODIO}

\author[1,2]{\fnm{Shi} \sur{Qianqian}}\email{shiqq@mail.sdu.edu.cn}
\author[1,2]{\fnm{Li} \sur{Teng}}\email{tengli@sdu.edu.cn}
\author*[1,2]{\fnm{Huang} \sur{Xingtao}}\email{huangxt@sdu.edu.cn}

\affil*[1]{\orgdiv{Institute of Frontier and Interdisciplinary Science}, \orgname{Shandong University}, \orgaddress{\street{72 Binhai Road}, \city{Qingdao}, \postcode{266237}, \state{Shandong}, \country{China}}}

\affil[2]{\orgdiv{Key Laboratory of Particle Physics and Particle Irradiation(MOE)}, \orgname{Shandong University}, \orgaddress{\street{72 Binhai Road}, \city{Qingdao}, \postcode{266237}, \state{Shandong}, \country{China}}}


\abstract{Software framework serves as a skeleton for the offline data processing software for many high energy physics (HEP) experiments. 
The event data management, including the event data model (EDM), transient event store and data input/output, implements the core functionalities of the framework, and has a great impact on the performance of the entire offline software. 
Future HEP experiments are generating increasingly large amounts of data, bringing challenges to offline data processing.
To address this issue, a common event data management system that supports efficient parallelized data processing applications has been developed based on SNiPER (Software for Non-collider Physics ExpeRiments) common software framework as well as PODIO, a common EDM toolkit for future HEP experiments. 
In this paper, the implementation of a parallelized event data management (PEDM) system is introduced, including the integration with MT-SNiPER and PODIO, as well as the implementation of GlobalStore to support multi-threaded event processing.
Finally, the application and performance evaluation of the data management system in OSCAR (offline software of Super Tau Charm Facility) is presented.}

\keywords{Offline data processing, SNiPER, high energy physics, event data model}

\maketitle

%%\linenumbers

\section{Introduction}
\label{chapter1}

Offline data processing software, such as Athena \citep{Calafiura:2005zz} for ATLAS \citep{atlas} and BOSS \citep{Li:2009ay} for BESIII \citep{BESIII}, is usually a critical component of modern high energy physics (HEP) experiments, responsible for building the offline data processing chain including detector simulation, calibration, reconstruction, and data analysis. 
The underlying framework, such as Gaudi \citep{Barrand:2001ny}, builds the foundation of the offline software, providing basic functionalities like event loop control, algorithm scheduling, event data management, detector description management, and some key common services.

The challenges in non-collider experiments such as nuclear reactor neutrino and cosmic ray experiments involve handling events with rare physics signals and correlation analysis between events within certain time windows, which are pretty different from collider experiments \citep{Yang}.
Learning the concepts of algorithm and service from Gaudi, and considering the needs of non-collider experiments, Chinese developers implemented the SNiPER framework \citep{Zou2015SNiPERAO}.
Now SNiPER has been adopted by several experiments, including Jiangmen Underground Neutrino Observatory (JUNO) \citep{Huang:2017dkh}, Large High Altitude Air Shower Observatory (LHAASO) \citep{Cao:2010zz}, Neutrinoless double beta decay experiment (nEXO) \citep{nEXO:2018ylp}.
SNiPER is very lightweight and easy-to-use.
It supports flexible event processing sequences and custom data management. 
As a general-purpose offline software framework, SNiPER is also suitable for collider experiments \citep{Li:2024tuy}, offering customizability, extensibility, and inherent advantages in parallel computing.
Therefore, we chose SNiPER as the underlying framework for the parallelized event data management (PEDM) system.

\begin{table}[htbp]
    \centering
    \caption{The amount of raw data generated by particle physics experiments each year}
    \begin{tabular}{ccc}
    \toprule
    Experiment  &Operation time  &Raw Data(PB) \\
    \midrule
    BESIII      &2008-2030       &0.2          \\
    LHAASO      &2018-2030       &9            \\
    JUNO        &2021-2030       &2            \\
    STCF        &--              &300          \\
    \bottomrule
    \end{tabular}
    \label{tab1}
\end{table}

The size of experimental data used for physics research is sharply increasing(see Table \ref{tab1}), which is a big challenge for offline data processing software. 
In order to take advantage of new computing resources, such as coprocessors, graphics processing units (GPUs), and field programmable gate array (FPGA), high-performance computing techniques must be considered and adopted to speedup data processing when designing and developing an offline software system for new generation experiments. 

\begin{figure}[htbp]
   \centering
   \includegraphics[width=1\linewidth]{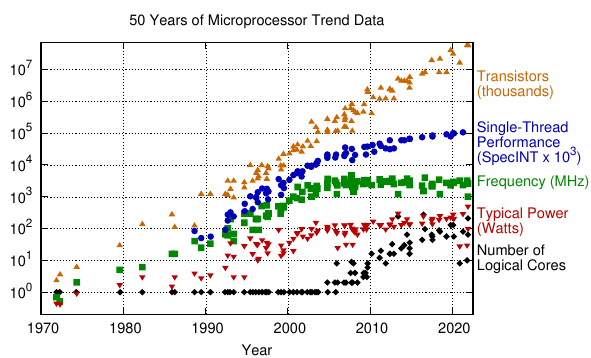}
   \caption{Microprocessor Trend Data since 1970 \citep{trend}. The figure shows that the amount of transistors increases year over year, but the growth rate of single-thread performance is gradually slowing down and approaching a plateau. Moreover, the rising number of CPU cores indicates the potential benefits of utilizing multi-core processing to enhance performance
   \label{fig:clock}}
\end{figure}

Recent developments in microprocessor technology have underlined the trends (Fig. \ref{fig:clock}) that have been identified since the mid-2000s: while the density of transistors on CPUs has continued to rise more or less exponentially (Moore’s Law \citep{658762}), clock speeds have plateaued.
The growth of single-threaded performance is gradually slowing down more than 15 years ago. 
To compensate for this, manufacturers put more cores on the chip. 
It is natural to apply parallel computing to take advantage of their high performance fully. 
In recent years, a multi-threading model has been adopted by the HEP community to develop several data processing frameworks and libraries.
For instance, CMS has already completed its transition to a concurrent framework \citep{Jones}.
GaudiHive \citep{Clemencic:2014cza}, an extension of Gaudi, has been a successful example in the HEP field, enabling concurrency through multi-threading.
In 2017, a multi-threaded version of SNiPER was developed, MT-SNiPER \citep{Zou:2018dqs}, which further enhanced processing performance.

Event data model (EDM) is the core of offline data processing software framework.
It defines the structure of event data in memory and data files and implements the relationship between data objects at different processing stages.
The PODIO toolkit \citep{Gaede:2021izq} provides an easy way to generate a performant implementation of an EDM from a high-level description in YAML format.
One of the key ideas of PODIO is to use plain-old-data (POD) type as much as feasible, encompassing structures composed of basic types, such as integers, floats and arrays of these basic types.
Thanks to their simplicity, efficiency, and ease of management, POD types enable parallelization. 
PODIO was chosen as the toolkit for the definition of EDM in the PEDM system.
Since POD types have simple internal structures and memory layouts,  they are highly suitable for sharing and transferring data between threads.

This paper aims to describe the PEDM system developed based on SNiPER and PODIO.
Section~\ref{chapter2} outlines the key components and the design of serial and parallelized event data management system.
Section~\ref{chapter3} provides a detailed description of the implementation of the PEDM system.
Section~\ref{chapter4} shows the performance of applying PEDM system to the offline software of Super Tau Charm Facility (OSCAR)~\citep{oscar}.
Finally, Section~\ref{chapter5} summarizes the status and gives an outlook for future development.

\section{Key Components and Design}
\label{chapter2}

In offline data processing software, the EDM is responsible for defining the structure of event data in memory and data files, and implementing the relationship between data objects in different processing stages.
The event data management system manages event data in memory, provides interfaces for user applications, and handles data input/output (I/O).
Therefore, the EDM and the event data management system greatly influence the function and performance of the experiment software.

\subsection{SNiPER Framework and PODIO}
\label{section1}

\begin{figure}[htbp]
   \centering
   \includegraphics[width=0.45\textwidth]{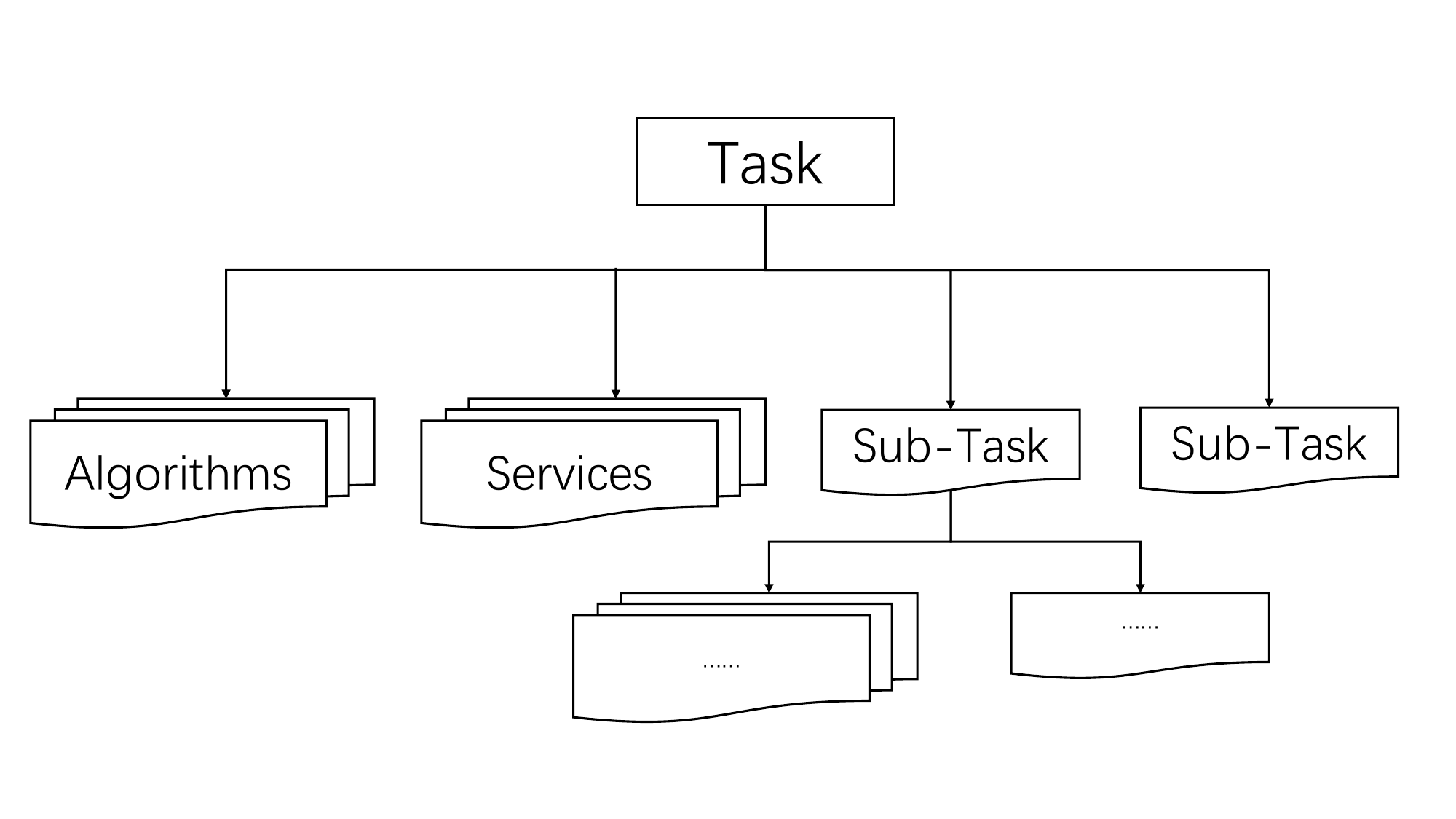}
   \caption{ Components tree in a SNiPER job. Each task comprises a set of Algorithms, Services, and Sub-Tasks
   \label{fig:task}}
\end{figure}

\begin{figure}
    \centering
    \includegraphics[width=0.42\textwidth]{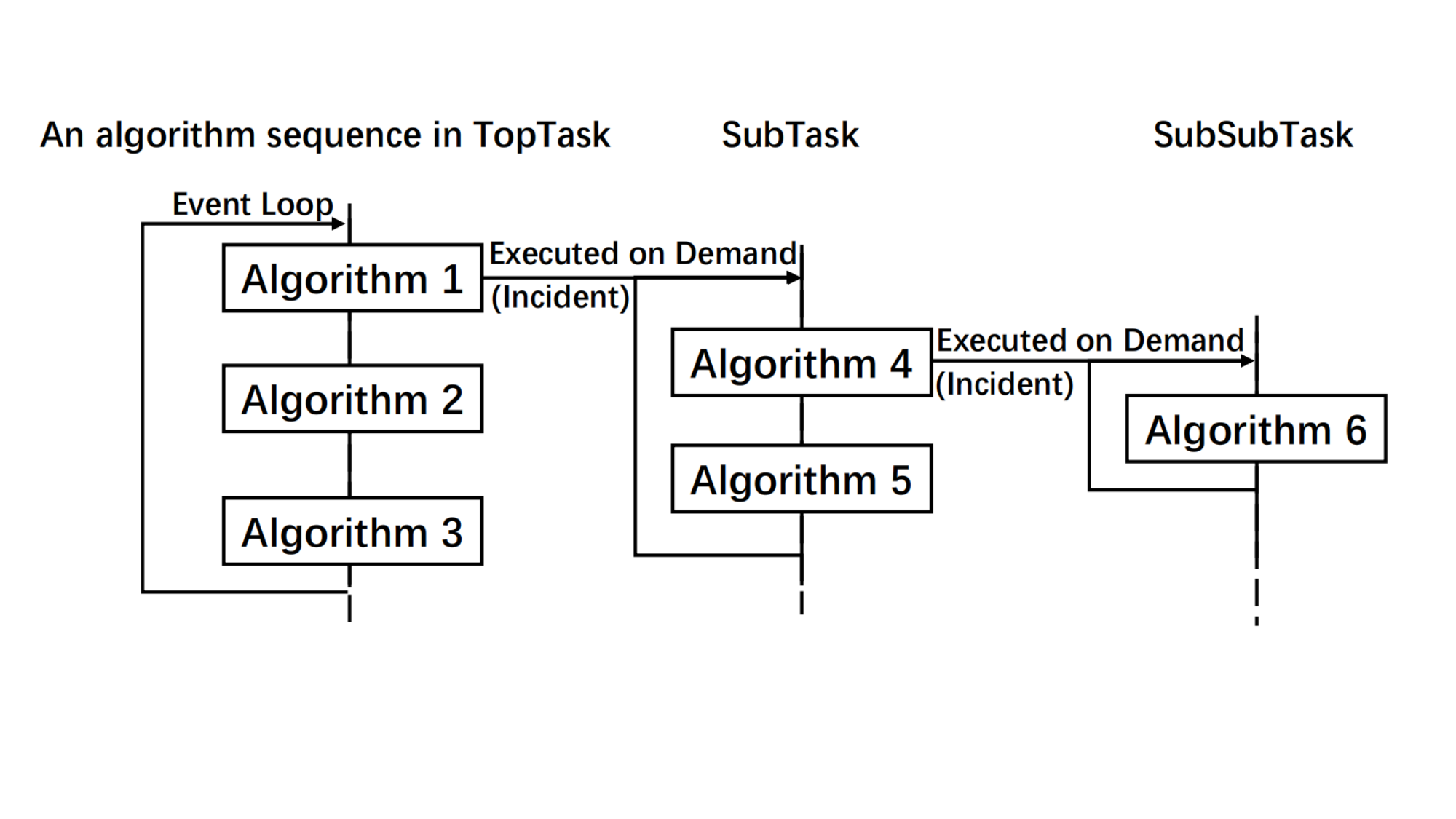}
    \caption{Conditional execution of algorithm subsets. Incident mechanism \citep{incidentSvc} is implemented to enhance communication between tasks. A task can be triggered by an incident on demand \label{fig:incident}}
\end{figure}

In the SNiPER framework, a parallelized solution based on Intel Threading Building Blocks (TBB) \citep{tbb}, MT-SNiPER, is proposed.
The \textit{Task} is one of the most essential components in SNiPER. 
It is responsible for controlling the event loop, functioning as the ``\textit{Application Manager}'' in Gaudi~\citep{Clemencic}.
As shown in Fig. \ref{fig:task}, each \textit{Task} contains a set of \textit{Algorithms}, \textit{Services}, and \textit{Sub-Tasks}.
Furthermore, SNiPER considers the need to use new parallel computing techniques in data processing.
It supports multiple \textit{Tasks}, which can be naturally mapped to threads in parallelized computing.
Fig. \ref{fig:incident} shows the execution sequence of multiple \textit{Tasks}, with each \textit{Task} managing its \textit{Algorithms} which are executed sequentially within a single \textit{Task}.

\begin{figure}[htbp]
   \centering
   \includegraphics[width=1\linewidth]{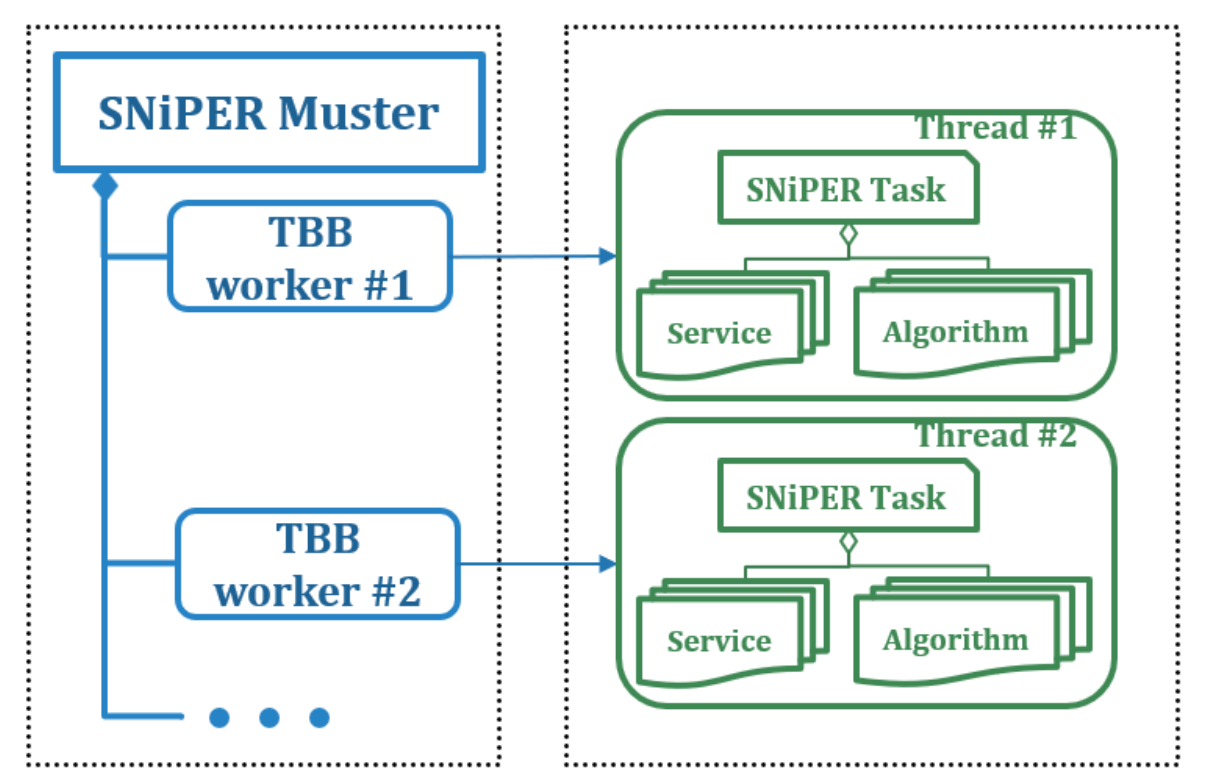}
   \caption{\label{fig:Muster}SNiPER Muster builds the mapping of a SNiPER Task to an Intel TBB task}
\end{figure}

Multiple SNiPER Task Scheduler (Muster) \citep{Zou:2019cyq} was implemented in MT-SNiPER to support multi-threading.
Muster builds the mapping of SNiPER \textit{Task} to Intel TBB task and breaks the traditional event loop.
Multiple instances of \textit{Tasks} are not executed directly by Muster; instead, Muster creates corresponding TBB-based workers to execute the \textit{Tasks}.
As shown in Fig. \ref{fig:Muster}, Muster spawns a set of workers and starts them in different threads. 
Then, different events are dispatched to different workers.
In each worker, it is a copy of the SNiPER \textit{Task} that performs like a serial job.
With this approach, the existing \textit{Algorithms} and \textit{Services} in serial mode can be almost seamlessly migrated to parallel mode.
  
\begin{figure}[htbp]
    \centering
    \includegraphics[width=1\linewidth]{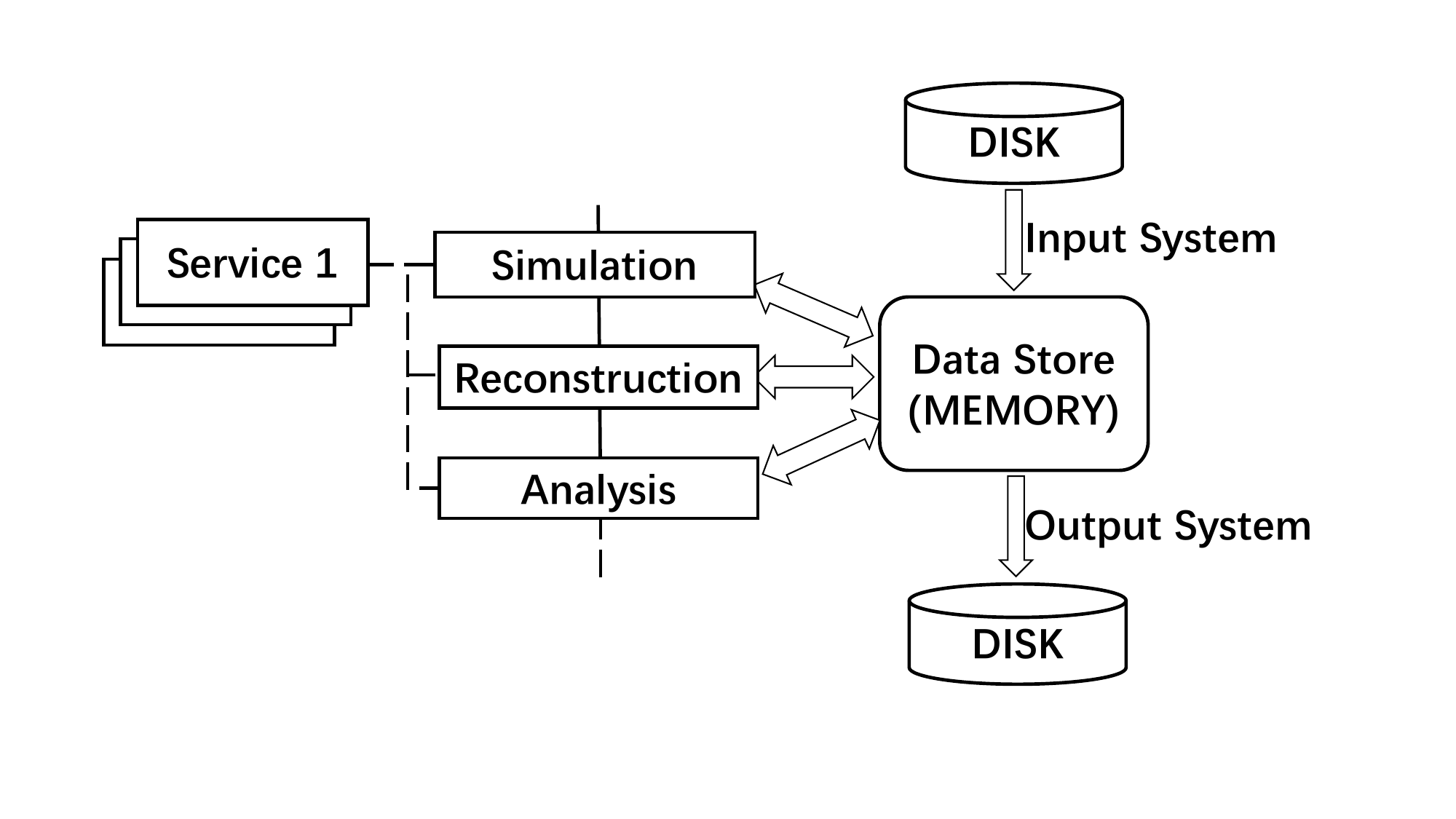}
    \caption{\label{fig:datastore}Diagram of the event data management system and its interfaces to applications}
\end{figure}

The event data management system of SNiPER is shown in Fig. \ref{fig:datastore}, including \textit{Input System}, \textit{Output System} and \textit{Data Store}.
\textit{Data Store} is responsible for handling event data in memory and managing its lifetime.
Each \textit{Task} instance has its own \textit{Data Store}.

As the \textit{Data Store} is used to manage general event data, we have chosen the \textit{EventStore} of PODIO to fulfill the functions of \textit{Data Store}.

The event data management system of PODIO mainly consists of three parts: the \textit{EventStore}, the \textit{ROOTReader}, and the \textit{ROOTWriter}. 
The \textit{ROOTReader} and \textit{ROOTWriter} manage data input and output. 
The \textit{EventStore} stores and manages the event data in memory and also provides the interfaces for \textit{ROOTReader} and \textit{ROOTWriter}.
By integrating the event data management systems of SNiPER with PODIO, an event data management system for serial data processing has been implemented in offline software.

Multi-threaded programming can effectively utilize multi-core CPU resources and allow different threads of the same task to run simultaneously. 
As mentioned earlier, Muster supports parallelized process control, but PODIO only holds data objects for one single event. 
It was necessary to redesign a \textit{EventStore} to cache multiple event data simultaneously.
Therefore, we designed the PEDM system by integrating Muster with PODIO to support parallelized data processing.

\subsection{Design of Serial Event Data Management System}
\label{section2}

\begin{figure*}[htbp]
   \begin{minipage}{0.25\textwidth}
     \caption{The sequence diagram for serial event data management. Tasks control the operation of SNiPER. The lifecycle of a Task covers the entire SNiPER job \label{fig:serial}}
   \end{minipage}
   \hfill
   \begin{minipage}{0.8\textwidth}
     \centering
     \includegraphics[width=1\linewidth]{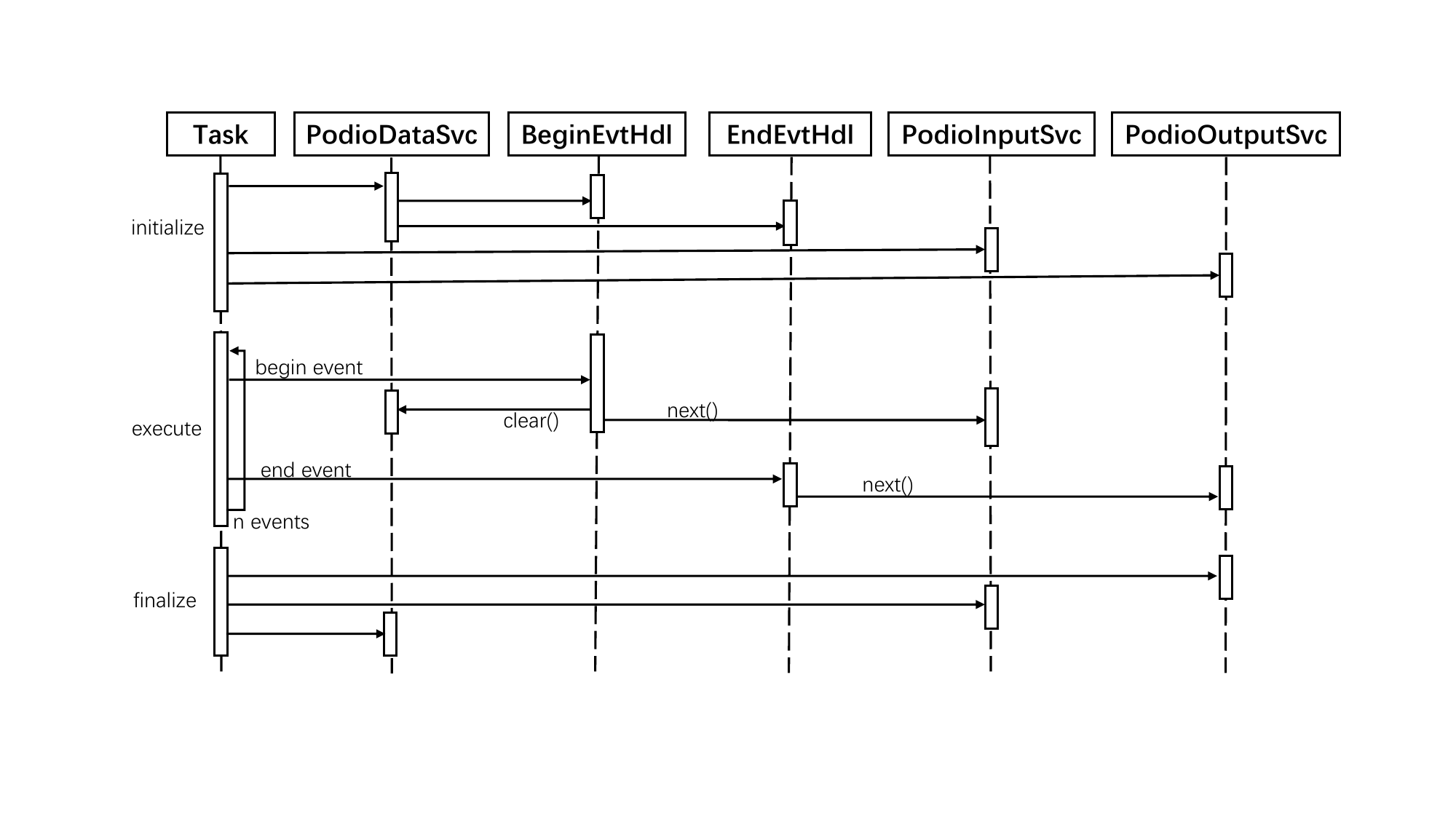}
   \end{minipage}
\end{figure*}

In the implementation of serial data processing, \textit{PodioDataSvc}, \textit{PodioInputSvc}, \textit{PodioOutputSvc} and \textit{DataHandle} are developed to integrate the PODIO with SNiPER.
The \textit{PodioDataSvc} serves as the wrapper of \textit{EventStore}, and there is only one \textit{EventStore} in a job.
The \textit{EventStore} in PODIO handles transient data, whose lifetime is managed by \textit{PodioDataSvc}.
Therefore, only one event can be processed at a time, and it needs to be promptly created and deleted to ensure the proper functioning of the program.
The sequence diagram for serial event data management is shown in Fig. \ref{fig:serial}.
In one task, a pair of incidents are named ``\textit{BeginEvtHdl}'' and ``\textit{EndEvtHdl}'' to implement the input and output of event data.
They are automatically triggered at the beginning and end of an event. 
Through the \textit{PodioDataSvc}, \textit{PodioInputSvc} is invoked by ``\textit{BeginEvtHdl}'' to convert persistent data from the input files to transient event data in the \textit{EventStore}, while \textit{PodioOutputSvc} is invoked by ``\textit{EndEvtHdl}'' to store the event data from the \textit{EventStore} into the output files.

\subsection{Design of Parallelized Event Data Management System}
\label{section3}

PODIO favors composition over inheritance and uses POD types to generate thread-safe and efficient C++ code for the EDM.
On the other hand, the \textit{EventStore} was never intended to support such use cases and has exceeded its original purpose as an example implementation for a transient event store \citep{podio}.
Therefore, we redesigned the \textit{EventStore} to support caching multiple events.

\begin{figure}[htbp]
   \centering
   \includegraphics[width=1\linewidth]{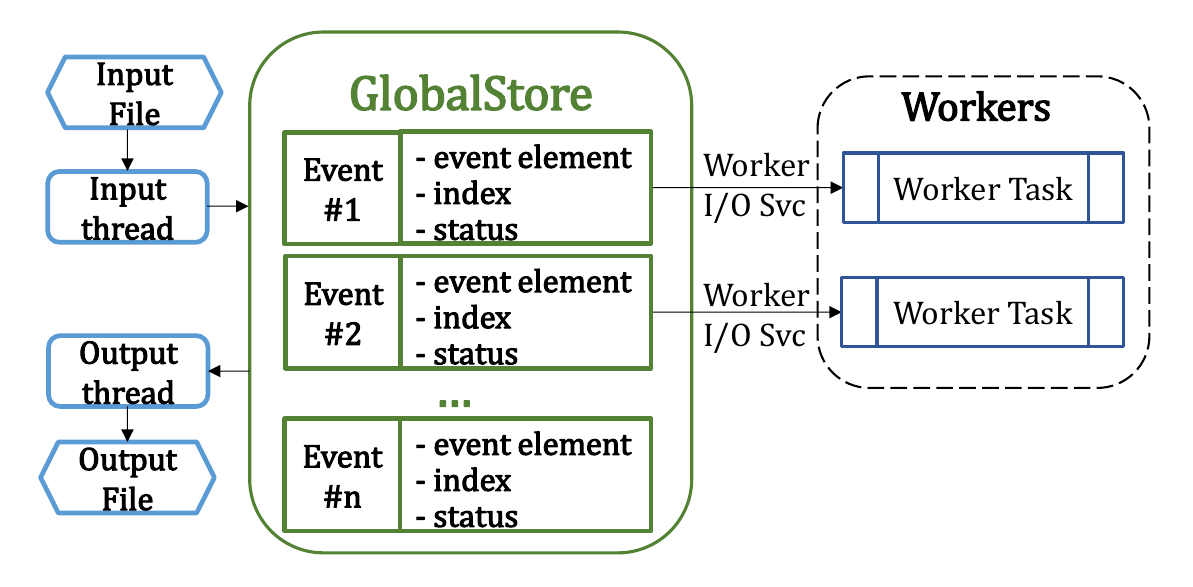}
   \caption{The design of event data management for parallel computing. Each Task can be configured with its own local buffer and corresponding input and output services \label{fig:GlobalStore}}
\end{figure}

Due to the flexibility of data management and the configuration of dynamic processes, it is critical to ensure memory safety.
Implementing security measures such as identity recognition, locking, and unlocking was essential to guarantee effective data processing and the consistency of data reading and writing.
While boosting the rate of offline data processing, this method ensures the reliability of results.
A new memory management system, named \textit{GlobalStore}(as illustrated in Fig. \ref{fig:GlobalStore}), has been designed and developed to support parallelized data processing fully.
In \textit{GlobalStore}, the structure named ``\textit{event element}'' is defined to manage most of the information that is handled by the \textit{EventStore}. 
\textit{GlobalStore} also stores the index and status to identify the event. 
The index is used to label the event, and the status is needed to ensure that the event is processed only once.

Data races arise when multiple threads simultaneously attempt to read/write data from/to the same file using ROOT I/O. 
To avoid this situation, decoupling the I/O functions from the \textit{EventStore} and allocating input and output to two dedicated threads is necessary.
This method significantly improves CPU utilization and reduces the task processing time, as described in Section \ref{chapter4}.

\begin{figure}[htbp]
   \centering
   \includegraphics[width=1\linewidth]{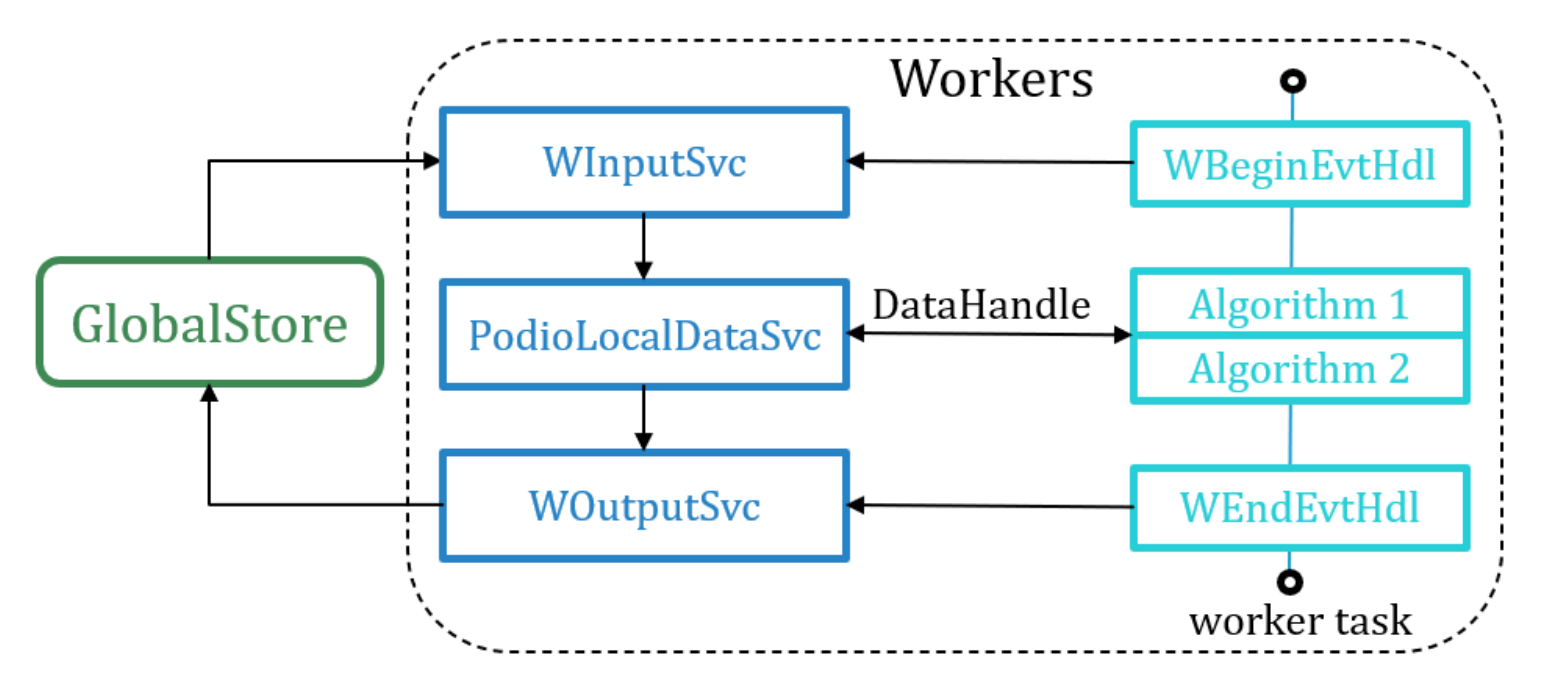}
   \caption{The data processing procedure of the worker. A worker is functioning as to a serial SNiPER application \label{fig:worker}}
\end{figure}

To efficiently deliver data to workers (see Fig. \ref{fig:worker}), the \textit{PodioLocalDataSvc} has been developed to access event data objects using \textit{DataHandles}. 
An abstract interface class, \textit{IPodioDataSvc}, has been defined to facilitate the sharing and reusing of a set of data management services between serial and parallelized event data management systems.
Both \textit{PodioDataSvc} and \textit{PodioLocalDataSvc} are inherited from \textit{IPodioDataSvc}, and these services are implemented to ensure thread safety.
The workers independently process events in parallel.
Two new I/O services, namely the worker input service (\textit{WInputSvc}) and the worker output service (\textit{WOutputSvc}), have been developed to deliver event data between \textit{GlobalStore} and workers.
They are triggered by two new incidents, ``\textit{WBeginEvtHdl}'' and ``\textit{WEndEvtHdl}''.
With this design, the intra-event and inter-event parallelism can be supported as described below.

\begin{itemize}
    \item The first technique employs thread synchronization mechanisms, including condition variables and mutex, to facilitate complex data exchange during data processing.
    After a worker thread acquires an event, it is locked by the mutex and released upon data processing completion, thus preventing a data race.
    Condition variables are used to coordinate synchronization between threads and ensure that data is not processed redundantly.
    Each worker is limited to the scope of a SNiPER Task.
    It is the same as a serial SNiPER job in most cases.
    
    \item The second technique, lazy loading, dynamically loads data to conserve memory and computational resources.
    Only the data that are accessed are loaded. 
    As a well-known technique, lazy loading has already been applied in some HEP frameworks \citep{MarlinMTP}.
    
\end{itemize}

Some options are offered to users for switching between serial and parallel modes based on their preferred job configuration.
Users have the freedom to define the number of worker threads.

\begin{figure*}[b]
   \centering
   \includegraphics[width=0.8\textwidth]{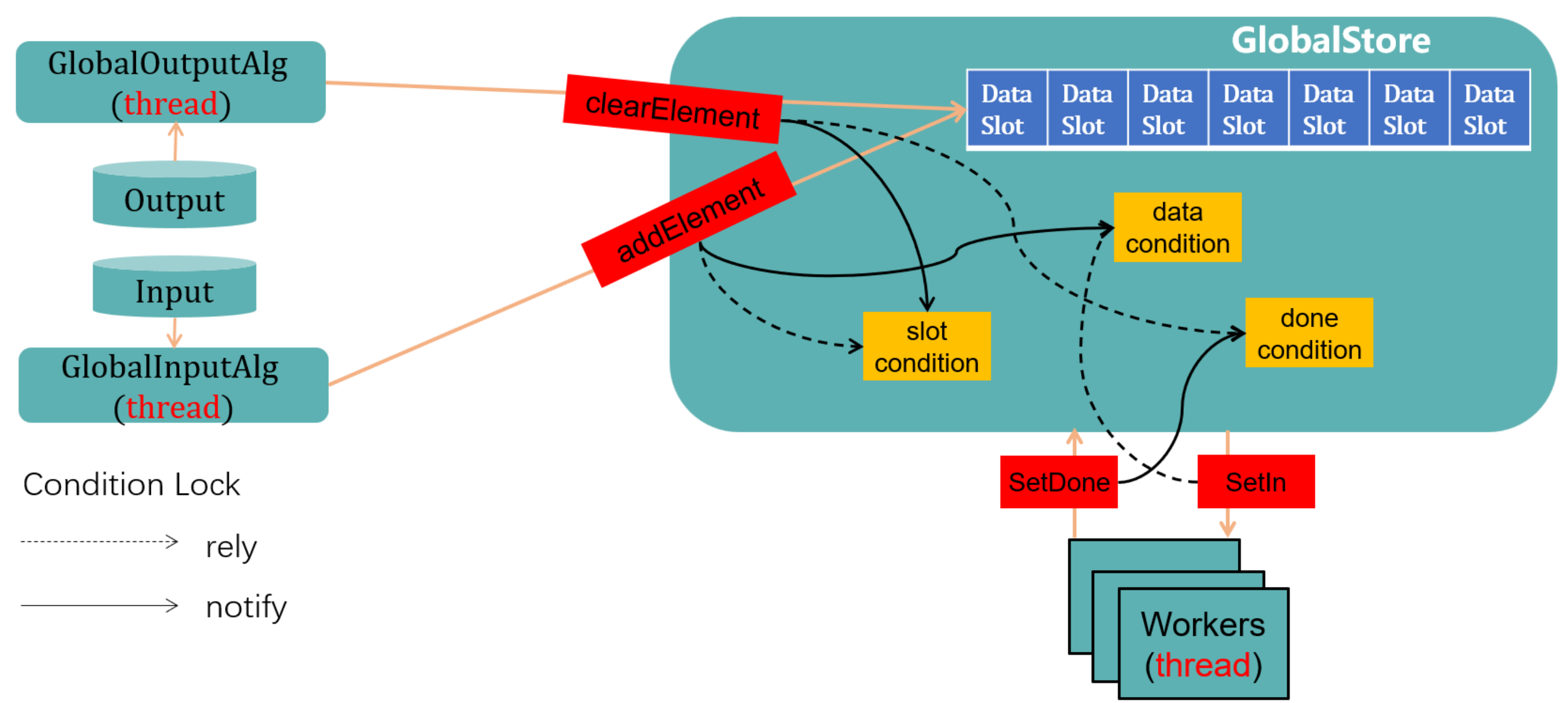}
   \caption{The implementation of the PEDM system \label{fig:overview}}
\end{figure*}

\section{Implementation of the Parallelized Event Data Management System}
\label{chapter3}

For caching more than one event, the \textit{GlobalStore} implements multiple data slots to store the ``\textit{event elements}'' as well as the status and ID corresponding to each event.
The information corresponding to a specific event can be easily obtained through its event ID, and each event comprises three states: ``\textit{Ready}'', ``\textit{Occupied}'', or ``\textit{Done}''. 
The states are utilized to guarantee that each event is processed exactly once.
Several condition variables are used in the \textit{GlobalStore} to ensure the safety of data exchange between threads.

The parallelized event data processing procedure is shown in Fig. \ref{fig:overview}.
Event data input and output are handled by two dedicated threads to achieve multi-threaded data processing, thereby improving CPU utilization.
The \textit{GlobalInputAlg} (input thread) reads event objects sequentially and sends them to \textit{GlobalStore}. 
The \textit{GlobalOutputAlg} (output thread) writes event objects into root files as persistent data when all events are completed. 

After an event is filled into the \textit{GlobalStore} by the input thread, it is marked as ``\textit{Ready}''. 
The primary role of the Muster scheduler is to create and manage worker threads.
As soon as the conditions are met, the worker thread immediately acquires and locks the event, marking it as ``\textit{Occupied}''. 

Customized I/O services have been implemented in the PEDM system to facilitate data exchange between the \textit{GlobalStore} and the workers. 
The input service gets event data directly from data slots, eliminating the necessity to read from a file.
Each worker is assigned to a thread-local SNiPER \textit{Task}, which is equivalent to a serial SNiPER job as described in Section~\ref{section2}.
In the local thread, we have developed the \textit{PodioLocalDataSvc} to handle the transient event data efficiently.
It utilizes a pair of incidents named \textit{WBeginEvtHdl} and \textit{WEndEvtHdl} to trigger the input and output services.
Additionally, the \textit{PodioLocalDataSvc} offers the capability to retrieve existing event data objects from memory or register a new event data object in memory.
This design enables efficient data management and manipulation within the worker.

\begin{figure*}[b]
   \centering
   \subcaptionbox{The distribution of reconstructed cluster energy \label{fig:energy}}{\includegraphics[width=0.3\textwidth]{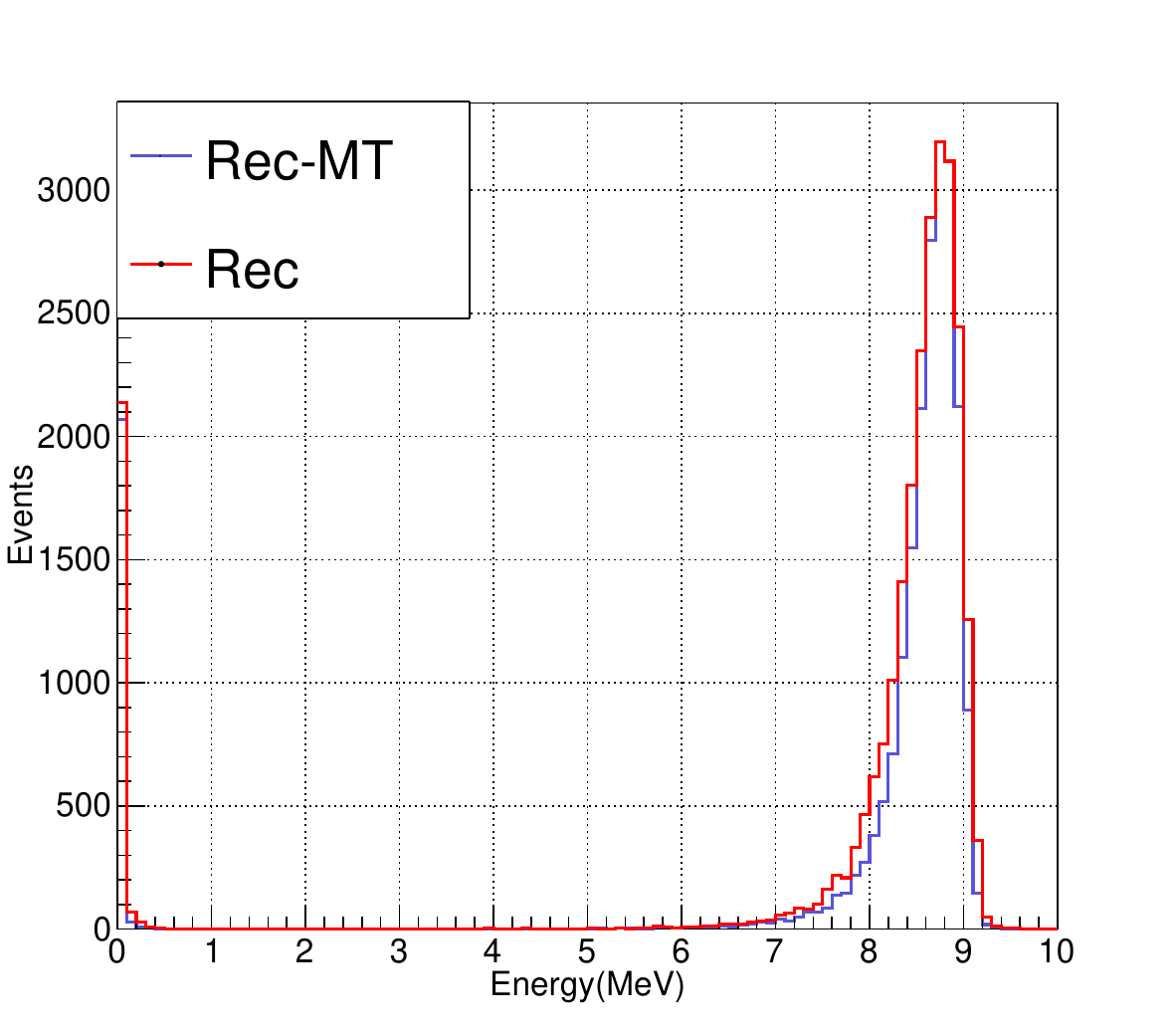}}
   \hfill
   \subcaptionbox{Energy deposition in the 3x3 crystals\label{fig:3x3}}{\includegraphics[width=0.3\textwidth]{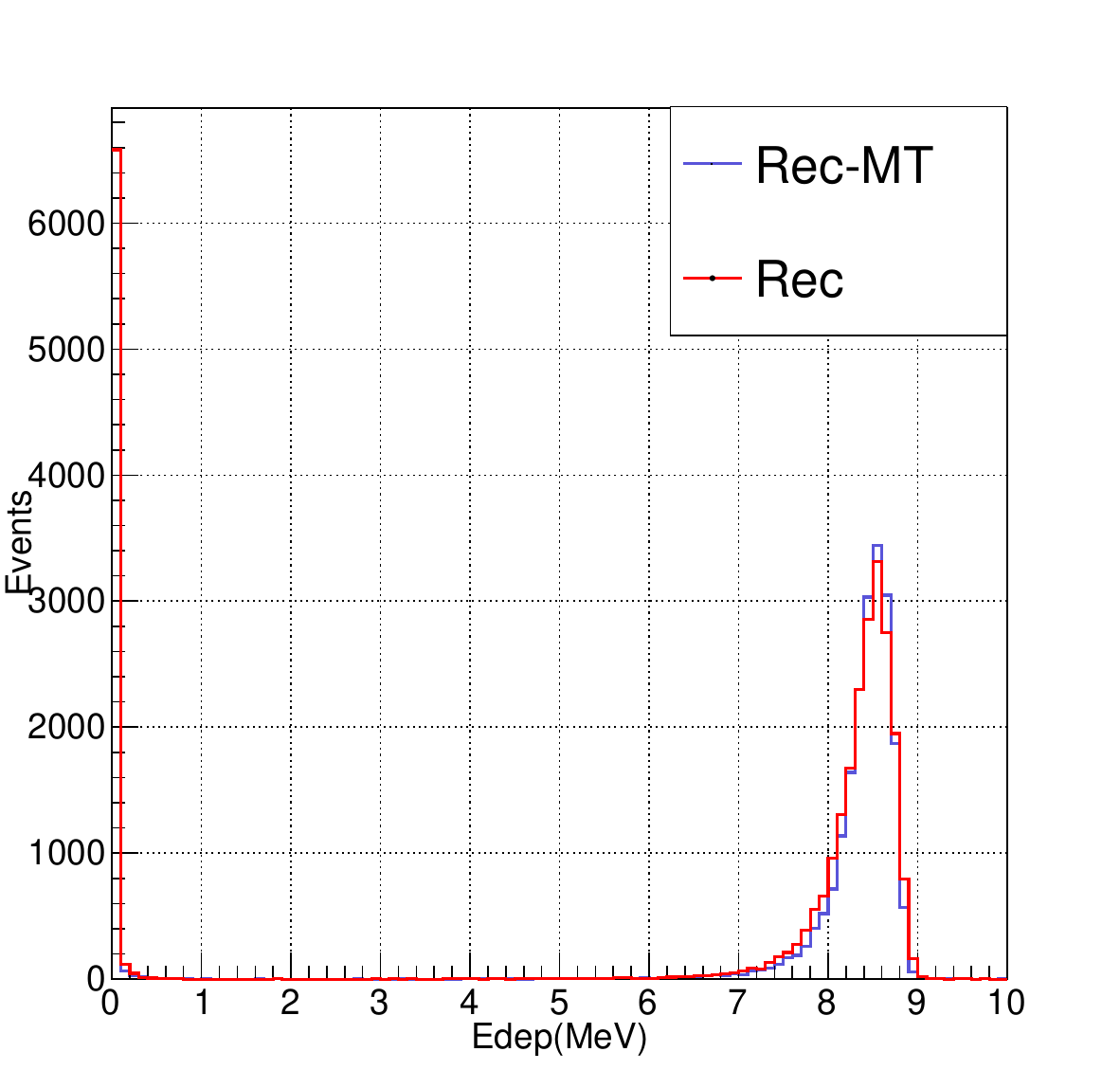}}
   \hfill
   \subcaptionbox{Energy deposition in the 5x5 crystals\label{fig:5x5}}{\includegraphics[width=0.3\textwidth]{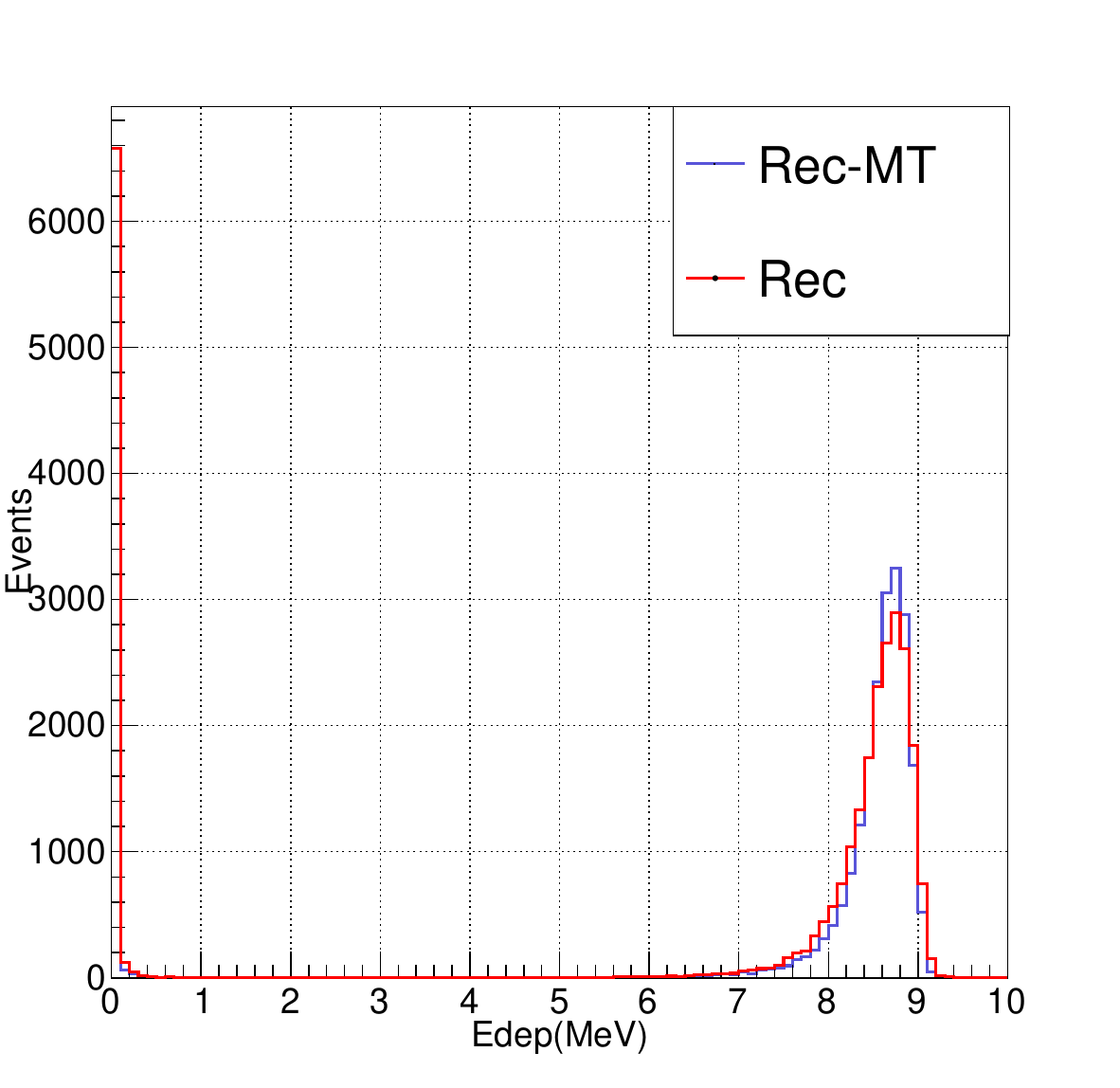}}
   \caption{The physics distribution of reconstruction results after simulating in two modes. The distributions are largely consistent between multi-threading (blue) and single-threading (red)\label{fig:result}}
   \label{fig:Result}
\end{figure*}

The algorithms responsible for processing event data are defined within the workers.
Once an event has been processed, the output service assigns it a ``\textit{Done}'' status without immediately writing it out. 
After all events have been processed, the output thread sequentially writes the marked events into the root file as persistent data. 
It is important to note that, to ensure optimal utilization of data slots, the \textit{GlobalStore} promptly clears these processed events to accommodate incoming events and prepare for the next event loop.

This implementation ensures that the processing of each event is entirely independent of other events. 
Events processed in worker threads can be written out without waiting for previously processed events in other threads, thereby reducing thread waiting time and improving CPU utilization.

\section{Application and Performance}
\label{chapter4}

The Super Tau Charm Facility (STCF) \citep{STCF} is a new-generation facility of electron-positron collider operating at center-of-mass energies of 2 to 7 GeV. 
STCF will play a leading role in the tau, charm, and hadron physics of HEP intensity frontier in the world.

The STCF will produce several hundreds of petabytes (PB) of scientific data annually.
To address this challenge, OSCAR is designed based on SNiPER \citep{ai} and partially based on Key4hep.
OSCAR is developed to facilitate the offline data processing tasks for the STCF experiment, including the production of Monte-Carlo simulation data, calibration, and reconstruction of collected data, as well as helping physicists to conduct physics analysis \citep{Huang_2023}.
The event data management system is a fundamental component for event data transfers and communications between OSCAR based applications in offline data processing.
Applying the PEDM system to OSCAR can further enhance the performance of OSCAR.

In our performance study, 20000 events of single e$^{-}$ with the energy of 5 GeV were generated at the collision point in STCF.
We executed full detector simulation and Electromagnetic calorimeter (EMC) reconstruction algorithms of different events in different threads and compared the reconstructed EMC information obtained in single-threaded mode with that obtained in multi-threaded mode to validate the functionality of the system.
The energy of reconstructed EMC clusters is the most crucial feature of EMC.
From Fig. \ref{fig:Result}, we can see that the energy distributions and the energy deposition distribution within the 3$\times$3 and 5$\times$5 crystals are largely consistent in both modes, demonstrating the effective functionality of the PEDM system.

\begin{figure}[htbp]
   \centering
   \includegraphics[width=0.45\textwidth]{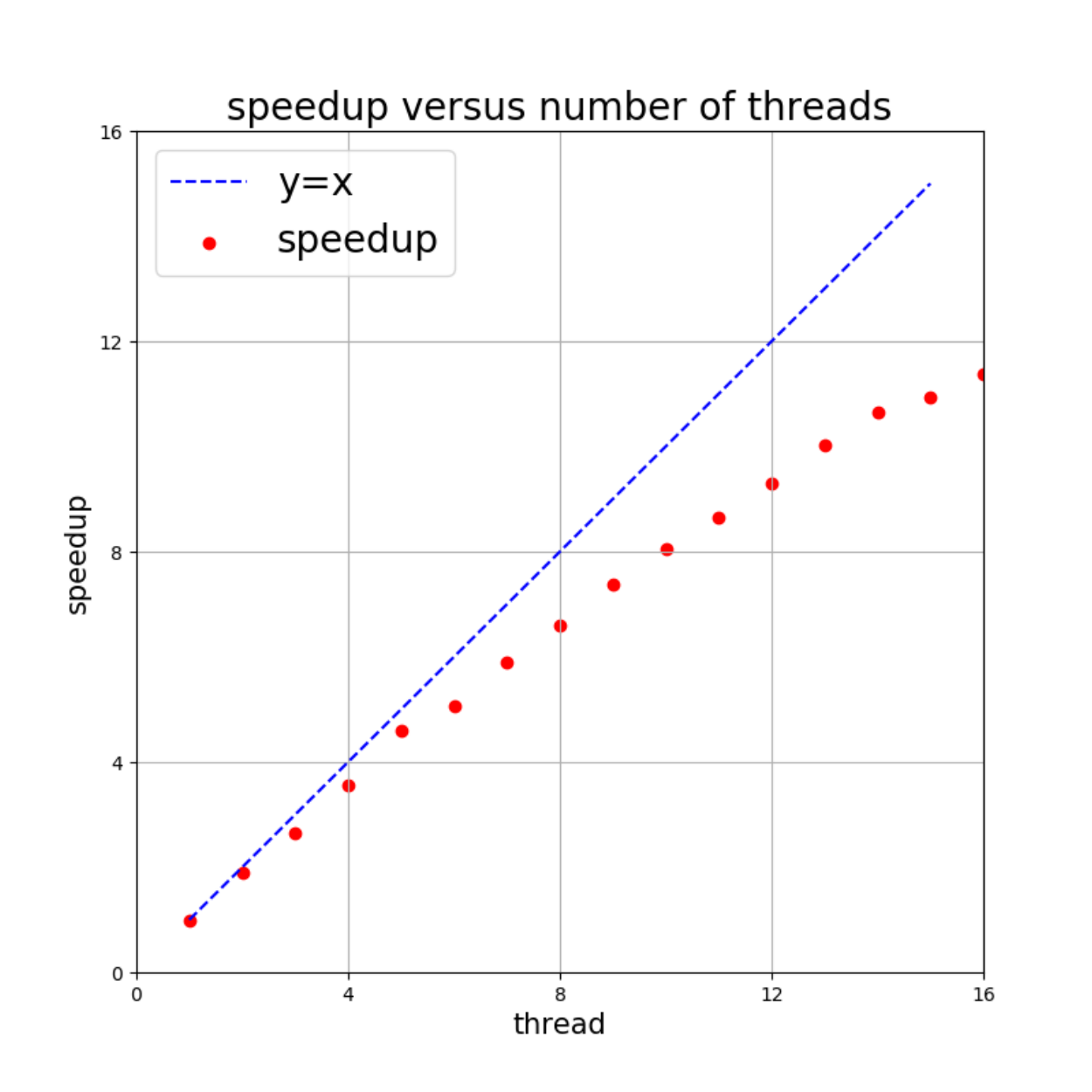}
   \caption{Speedup ratio versus number of worker threads. The speedup ratio is close to the ideal value with less than 5 threads \label{fig:performance}}
\end{figure}

To measure the speedup, each job was repeated three times to calculate the average time and then the average time was divided by the serial time.
The speedup ratio is plotted against the number of threads in Fig. \ref{fig:performance}.
It can be observed that when the number of threads is less than 5, the speedup ratio exhibits a good linear behavior.
However, beyond that point, factors such as thread scheduling cause the speedup ratio to gradually deviate from the ideal line.
This is because the workload per task is not substantial enough.
Our performance study indicates that the PEDM system can reduce data processing time and meet the requirements to handle large amounts of data. 

\section{Conclusions}
\label{chapter5}

SNiPER is a simple and lightweight software framework that has been used in several HEP experiments.
To meet the requirements of parallel computing, a common event data management system is redesigned by consistently integrating MT-SNiPER and PODIO.
Furthermore, the PEDM system has been designed and developed to support parallelized event data management and processing well.
Its application and performance study in OSCAR has shown that the PEDM system speeds up the event data processing with the advantage of concurrent event processing.
At the same time, due to its good generality and flexibility in the design and implementation, other HEP experiments can also easily adopt it to implement their parallel data processing.

\backmatter

\bmhead{Acknowledgements}

This work was performed with the support of the National Natural Science Foundation of China (No.: 12025502, 12341504, 12105158), National Key Research and Development Program of China (Grant No.: 2021YFA0718403).

\bibliography{sn-bibliography}% common bib file


\begin{thebibliography}{27}
\ifx \bisbn   \undefined \def \bisbn  #1{ISBN #1}\fi
\ifx \binits  \undefined \def \binits#1{#1}\fi
\ifx \bauthor  \undefined \def \bauthor#1{#1}\fi
\ifx \batitle  \undefined \def \batitle#1{#1}\fi
\ifx \bjtitle  \undefined \def \bjtitle#1{#1}\fi
\ifx \bvolume  \undefined \def \bvolume#1{\textbf{#1}}\fi
\ifx \byear  \undefined \def \byear#1{#1}\fi
\ifx \bissue  \undefined \def \bissue#1{#1}\fi
\ifx \bfpage  \undefined \def \bfpage#1{#1}\fi
\ifx \blpage  \undefined \def \blpage #1{#1}\fi
\ifx \burl  \undefined \def \burl#1{\textsf{#1}}\fi
\ifx \doiurl  \undefined \def \doiurl#1{\url{https://doi.org/#1}}\fi
\ifx \betal  \undefined \def \betal{\textit{et al.}}\fi
\ifx \binstitute  \undefined \def \binstitute#1{#1}\fi
\ifx \binstitutionaled  \undefined \def \binstitutionaled#1{#1}\fi
\ifx \bctitle  \undefined \def \bctitle#1{#1}\fi
\ifx \beditor  \undefined \def \beditor#1{#1}\fi
\ifx \bpublisher  \undefined \def \bpublisher#1{#1}\fi
\ifx \bbtitle  \undefined \def \bbtitle#1{#1}\fi
\ifx \bedition  \undefined \def \bedition#1{#1}\fi
\ifx \bseriesno  \undefined \def \bseriesno#1{#1}\fi
\ifx \blocation  \undefined \def \blocation#1{#1}\fi
\ifx \bsertitle  \undefined \def \bsertitle#1{#1}\fi
\ifx \bsnm \undefined \def \bsnm#1{#1}\fi
\ifx \bsuffix \undefined \def \bsuffix#1{#1}\fi
\ifx \bparticle \undefined \def \bparticle#1{#1}\fi
\ifx \barticle \undefined \def \barticle#1{#1}\fi
\bibcommenthead
\ifx \bconfdate \undefined \def \bconfdate #1{#1}\fi
\ifx \botherref \undefined \def \botherref #1{#1}\fi
\ifx \url \undefined \def \url#1{\textsf{#1}}\fi
\ifx \bchapter \undefined \def \bchapter#1{#1}\fi
\ifx \bbook \undefined \def \bbook#1{#1}\fi
\ifx \bcomment \undefined \def \bcomment#1{#1}\fi
\ifx \oauthor \undefined \def \oauthor#1{#1}\fi
\ifx \citeauthoryear \undefined \def \citeauthoryear#1{#1}\fi
\ifx \endbibitem  \undefined \def \endbibitem {}\fi
\ifx \bconflocation  \undefined \def \bconflocation#1{#1}\fi
\ifx \arxivurl  \undefined \def \arxivurl#1{\textsf{#1}}\fi
\csname PreBibitemsHook\endcsname

%%% 1
\bibitem[\protect\citeauthoryear{Calafiura et~al.}{2005}]{Calafiura:2005zz}
\begin{bchapter}
\bauthor{\bsnm{Calafiura}, \binits{P.}},
\bauthor{\bsnm{Marino}, \binits{M.}},
\bauthor{\bsnm{Leggett}, \binits{C.}},
\bauthor{\bsnm{Lavrijsen}, \binits{W.}},
\bauthor{\bsnm{Quarrie}, \binits{D.R.}}:
\bctitle{The athena control framework in production, new developments and lessons learned}.
In: \bbtitle{14th International Conference on Computing in High-Energy and Nuclear Physics},
pp. \bfpage{456}--\blpage{458}
(\byear{2005}).
\burl{https://api.semanticscholar.org/CorpusID:73681202}
\end{bchapter}
\endbibitem

%%% 2
\bibitem[\protect\citeauthoryear{Aad et~al.}{2008}]{atlas}
\begin{barticle}
\bauthor{\bsnm{Aad}, \binits{G.}},
\bauthor{\bsnm{Abat}, \binits{E.}},
\bauthor{\bsnm{Abdallah}, \binits{J.}}, \betal:
\batitle{The atlas experiment at the cern large hadron collider}.
\bjtitle{Journal of Instrumentation}
\bvolume{3}(\bissue{08}),
\bfpage{08003}
(\byear{2008})
\doiurl{10.1088/1748-0221/3/08/S08003}
\end{barticle}
\endbibitem

%%% 3
\bibitem[\protect\citeauthoryear{Li et~al.}{2009}]{Li:2009ay}
\begin{barticle}
\bauthor{\bsnm{Li}, \binits{W.-D.}},
\bauthor{\bsnm{Mao}, \binits{Y.-J.}},
\bauthor{\bsnm{Wang}, \binits{Y.-F.}}:
\batitle{The bes-iii detector and offline software}.
\bjtitle{Int. J. Mod. Phys. A}
\bvolume{24S1},
\bfpage{9}--\blpage{21}
(\byear{2009})
\doiurl{10.1142/S0217751X09046424}
\end{barticle}
\endbibitem

%%% 4
\bibitem[\protect\citeauthoryear{Ablikim et~al.}{2010}]{BESIII}
\begin{barticle}
\bauthor{\bsnm{Ablikim}, \binits{M.}},
\bauthor{\bsnm{An}, \binits{Z.H.}}, \betal:
\batitle{Design and construction of the besiii detector}.
\bjtitle{Nuclear Instruments and Methods in Physics Research Section A: Accelerators, Spectrometers, Detectors and Associated Equipment}
\bvolume{614}(\bissue{3}),
\bfpage{345}--\blpage{399}
(\byear{2010})
\doiurl{10.1016/j.nima.2009.12.050}
\end{barticle}
\endbibitem

%%% 5
\bibitem[\protect\citeauthoryear{Barrand et~al.}{2001}]{Barrand:2001ny}
\begin{barticle}
\bauthor{\bsnm{Barrand}, \binits{G.}}, \betal:
\batitle{Gaudi - a software architecture and framework for building hep data processing applications}.
\bjtitle{Comput. Phys. Commun.}
\bvolume{140},
\bfpage{45}--\blpage{55}
(\byear{2001})
\doiurl{10.1016/S0010-4655(01)00254-5}
\end{barticle}
\endbibitem

%%% 6
\bibitem[\protect\citeauthoryear{Yang and on~behalf of~the JUNO~collaboration}{2023}]{Yang}
\begin{barticle}
\bauthor{\bsnm{Yang}, \binits{Y.}},
\bauthor{\bsnm{JUNO~collaboration}}:
\batitle{Parallel processing in data analysis of the juno experiment}.
\bjtitle{Journal of Physics: Conference Series}
\bvolume{2438}(\bissue{1}),
\bfpage{012057}
(\byear{2023})
\doiurl{10.1088/1742-6596/2438/1/012057}
\end{barticle}
\endbibitem

%%% 7
\bibitem[\protect\citeauthoryear{Zou et~al.}{2015}]{Zou2015SNiPERAO}
\begin{barticle}
\bauthor{\bsnm{Zou}, \binits{J.H.}},
\bauthor{\bsnm{Huang}, \binits{X.}},
\bauthor{\bsnm{Li}, \binits{W.}},
\bauthor{\bsnm{Lin}, \binits{T.}},
\bauthor{\bsnm{Li}, \binits{T.}},
\bauthor{\bsnm{Zhang}, \binits{K.}},
\bauthor{\bsnm{Deng}, \binits{Z.Y.}},
\bauthor{\bsnm{Cao}, \binits{G.F.}}:
\batitle{Sniper: an offline software framework for non-collider physics experiments}.
\bjtitle{Journal of Physics: Conference Series}
\bvolume{664}(\bissue{7}),
\bfpage{072053}
(\byear{2015})
\doiurl{10.1088/1742-6596/664/7/072053}
\end{barticle}
\endbibitem

%%% 8
\bibitem[\protect\citeauthoryear{Huang et~al.}{2017}]{Huang:2017dkh}
\begin{barticle}
\bauthor{\bsnm{Huang}, \binits{X.}},
\bauthor{\bsnm{Li}, \binits{T.}},
\bauthor{\bsnm{Zou}, \binits{J.}},
\bauthor{\bsnm{Lin}, \binits{T.}},
\bauthor{\bsnm{Li}, \binits{W.}},
\bauthor{\bsnm{Deng}, \binits{Z.}},
\bauthor{\bsnm{Cao}, \binits{G.}}:
\batitle{Offline data processing software for the juno experiment}.
\bjtitle{PoS}
\bvolume{ICHEP2016},
\bfpage{1051}
(\byear{2017})
\doiurl{10.22323/1.282.1051}
\end{barticle}
\endbibitem

%%% 9
\bibitem[\protect\citeauthoryear{Cao}{2010}]{Cao:2010zz}
\begin{barticle}
\bauthor{\bsnm{Cao}, \binits{Z.}}:
\batitle{A future project at tibet: The large high altitude air shower observatory (lhaaso)}.
\bjtitle{Chin. Phys. C}
\bvolume{34},
\bfpage{249}--\blpage{252}
(\byear{2010})
\doiurl{10.1088/1674-1137/34/2/018}
\end{barticle}
\endbibitem

%%% 10
\bibitem[\protect\citeauthoryear{Kharusi et~al.}{2018}]{nEXO:2018ylp}
\begin{botherref}
\oauthor{\bsnm{Kharusi}, \binits{S.A.}}, et al.:
nEXO Pre-Conceptual Design Report
(2018)
\end{botherref}
\endbibitem

%%% 11
\bibitem[\protect\citeauthoryear{Li et~al.}{2024}]{Li:2024tuy}
\begin{barticle}
\bauthor{\bsnm{Li}, \binits{T.}},
\bauthor{\bsnm{Huang}, \binits{W.}},
\bauthor{\bsnm{Huang}, \binits{X.}},
\bauthor{\bsnm{Ai}, \binits{X.}},
\bauthor{\bsnm{Li}, \binits{H.}},
\bauthor{\bsnm{Liu}, \binits{D.}}:
\batitle{Offline data processing software for the super tau charm facility}.
\bjtitle{EPJ Web Conf.}
\bvolume{295},
\bfpage{03025}
(\byear{2024})
\doiurl{10.1051/epjconf/202429503025}
\end{barticle}
\endbibitem

%%% 12
\bibitem[\protect\citeauthoryear{Moore}{1998}]{658762}
\begin{barticle}
\bauthor{\bsnm{Moore}, \binits{G.E.}}:
\batitle{Cramming more components onto integrated circuits}.
\bjtitle{Proceedings of the IEEE}
\bvolume{86}(\bissue{1}),
\bfpage{82}--\blpage{85}
(\byear{1998})
\doiurl{10.1109/JPROC.1998.658762}
\end{barticle}
\endbibitem

%%% 13
\bibitem[\protect\citeauthoryear{Jones and Sexton-Kennedy}{2014}]{Jones}
\begin{barticle}
\bauthor{\bsnm{Jones}, \binits{C.D.}},
\bauthor{\bsnm{Sexton-Kennedy}, \binits{E.}}:
\batitle{Stitched together: Transitioning cms to a hierarchical threaded framework}.
\bjtitle{Journal of Physics: Conference Series}
\bvolume{513}(\bissue{2}),
\bfpage{022034}
(\byear{2014})
\doiurl{10.1088/1742-6596/513/2/022034}
\end{barticle}
\endbibitem

%%% 14
\bibitem[\protect\citeauthoryear{Clemencic et~al.}{2014}]{Clemencic:2014cza}
\begin{barticle}
\bauthor{\bsnm{Clemencic}, \binits{M.}},
\bauthor{\bsnm{Hegner}, \binits{B.}},
\bauthor{\bsnm{Mato}, \binits{P.}},
\bauthor{\bsnm{Piparo}, \binits{D.}}:
\batitle{Introducing concurrency in the gaudi data processing framework}.
\bjtitle{J. Phys. Conf. Ser.}
\bvolume{513},
\bfpage{022013}
(\byear{2014})
\doiurl{10.1088/1742-6596/513/2/022013}
\end{barticle}
\endbibitem

%%% 15
\bibitem[\protect\citeauthoryear{Zou et~al.}{2018}]{Zou:2018dqs}
\begin{barticle}
\bauthor{\bsnm{Zou}, \binits{J.H.}},
\bauthor{\bsnm{Lin}, \binits{T.}},
\bauthor{\bsnm{Li}, \binits{W.D.}},
\bauthor{\bsnm{Huang}, \binits{X.T.}},
\bauthor{\bsnm{Li}, \binits{T.}},
\bauthor{\bsnm{Deng}, \binits{Z.Y.}},
\bauthor{\bsnm{Cao}, \binits{G.F.}},
\bauthor{\bsnm{You}, \binits{Z.Y.}}:
\batitle{Parallel computing of sniper based on intel tbb}.
\bjtitle{J. Phys. Conf. Ser.}
\bvolume{1085}(\bissue{3}),
\bfpage{032009}
(\byear{2018})
\doiurl{10.1088/1742-6596/1085/3/032009}
\end{barticle}
\endbibitem

%%% 16
\bibitem[\protect\citeauthoryear{{Karl Rupp}}{}]{trend}
\begin{botherref}
\oauthor{\bsnm{{Karl Rupp}}}:
microprocessor trend data.
\url{https://github.com/karlrupp/microprocessor-trend-data/tree/master/50yrs}
\end{botherref}
\endbibitem

%%% 17
\bibitem[\protect\citeauthoryear{Gaede et~al.}{2021}]{Gaede:2021izq}
\begin{barticle}
\bauthor{\bsnm{Gaede}, \binits{F.}},
\bauthor{\bsnm{Ganis}, \binits{G.}},
\bauthor{\bsnm{Hegner}, \binits{B.}},
\bauthor{\bsnm{Helsens}, \binits{C.}},
\bauthor{\bsnm{Madlener}, \binits{T.}},
\bauthor{\bsnm{Sailer}, \binits{A.}},
\bauthor{\bsnm{Stewart}, \binits{G.A.}},
\bauthor{\bsnm{Volkl}, \binits{V.}},
\bauthor{\bsnm{Wang}, \binits{J.}}:
\batitle{Edm4hep and podio - the event data model of the key4hep project and its implementation}.
\bjtitle{EPJ Web Conf.}
\bvolume{251},
\bfpage{03026}
(\byear{2021})
\doiurl{10.1051/epjconf/202125103026}
\end{barticle}
\endbibitem

%%% 18
\bibitem[\protect\citeauthoryear{Huang et~al.}{2023}]{oscar}
\begin{barticle}
\bauthor{\bsnm{Huang}, \binits{W.H.}},
\bauthor{\bsnm{Li}, \binits{H.}},
\bauthor{\bsnm{Zhou}, \binits{H.}},
\bauthor{\bsnm{Li}, \binits{T.}},
\bauthor{\bsnm{Li}, \binits{Q.Y.}},
\bauthor{\bsnm{Huang}, \binits{X.T.}}:
\batitle{Design and development of the core software for stcf offline data processing}.
\bjtitle{JINST}
\bvolume{18}(\bissue{03}),
\bfpage{03004}
(\byear{2023})
\doiurl{10.1088/1748-0221/18/03/P03004}
{\href{https://arxiv.org/abs/2211.03137}{{arXiv:2211.03137}}}
{[physics.ins-det]}
\end{barticle}
\endbibitem

%%% 19
\bibitem[\protect\citeauthoryear{Robison}{2011}]{tbb}
\begin{bbook}
\bauthor{\bsnm{Robison}, \binits{A.D.}}:
In: \beditor{\bsnm{Padua}, \binits{D.}} (ed.)
\bbtitle{Intel® Threading Building Blocks (TBB)},
pp. \bfpage{955}--\blpage{964}.
\bpublisher{Springer},
\blocation{Boston, MA}
(\byear{2011}).
\doiurl{10.1007/978-0-387-09766-4_51}
\end{bbook}
\endbibitem

%%% 20
\bibitem[\protect\citeauthoryear{Clemencic et~al.}{2011}]{Clemencic}
\begin{barticle}
\bauthor{\bsnm{Clemencic}, \binits{M.}},
\bauthor{\bsnm{Corti}, \binits{G.}},
\bauthor{\bsnm{Easo}, \binits{S.}},
\bauthor{\bsnm{Jones}, \binits{C.R.}},
\bauthor{\bsnm{Miglioranzi}, \binits{S.}},
\bauthor{\bsnm{Pappagallo}, \binits{M.}},
\bauthor{\bsnm{Robbe}, \binits{P.}}:
\batitle{The lhcb simulation application, gauss: Design, evolution and experience}.
\bjtitle{J. Phys. Conf. Ser.}
\bvolume{331},
\bfpage{032023}
(\byear{2011})
\doiurl{10.1088/1742-6596/331/3/032023}
\end{barticle}
\endbibitem

%%% 21
\bibitem[\protect\citeauthoryear{}{}]{incidentSvc}
\begin{botherref}
Gaudi framework services.
\url{https://gaudi-framework.readthedocs.io/en/latest/old/GDG_Services.html}
\end{botherref}
\endbibitem

%%% 22
\bibitem[\protect\citeauthoryear{Zou et~al.}{2019}]{Zou:2019cyq}
\begin{barticle}
\bauthor{\bsnm{Zou}, \binits{J.}},
\bauthor{\bsnm{Lin}, \binits{T.}},
\bauthor{\bsnm{Li}, \binits{W.}},
\bauthor{\bsnm{Huang}, \binits{X.}},
\bauthor{\bsnm{Deng}, \binits{Z.}},
\bauthor{\bsnm{Cao}, \binits{G.}},
\bauthor{\bsnm{You}, \binits{Z.}}:
\batitle{The event buffer management for mt-sniper}.
\bjtitle{EPJ Web Conf.}
\bvolume{214},
\bfpage{05026}
(\byear{2019})
\doiurl{10.1051/epjconf/201921405026}
\end{barticle}
\endbibitem

%%% 23
\bibitem[\protect\citeauthoryear{Fernandez~Declara et~al.}{2023}]{podio}
\begin{bchapter}
\bauthor{\bsnm{Fernandez~Declara}, \binits{P.}},
\bauthor{\bsnm{Gaede}, \binits{F.}},
\bauthor{\bsnm{Ganis}, \binits{G.}},
\bauthor{\bsnm{Hegner}, \binits{B.}},
\bauthor{\bsnm{Helsens}, \binits{C.}},
\bauthor{\bsnm{Madlener}, \binits{T.}},
\bauthor{\bsnm{Sailer}, \binits{A.}},
\bauthor{\bsnm{Stewart}, \binits{G.A.}},
\bauthor{\bsnm{Volkl}, \binits{V.}}:
\bctitle{Of frames and schema evolution -- the newest features of podio}.
In: \bbtitle{21th International Workshop on Advanced Computing and Analysis Techniques in Physics Research: AI Meets Reality}
(\byear{2023}).
\burl{https://arxiv.org/abs/2312.08199}
\end{bchapter}
\endbibitem

%%% 24
\bibitem[\protect\citeauthoryear{Et{\'e} et~al.}{2020}]{MarlinMTP}
\begin{botherref}
\oauthor{\bsnm{Et{\'e}}, \binits{R.}},
\oauthor{\bsnm{Gaede}, \binits{F.}},
\oauthor{\bsnm{Benda}, \binits{J.}},
\oauthor{\bsnm{Grasland}, \binits{H.}}:
Marlinmt - parallelising the marlin framework.
EPJ Web of Conferences
(2020)
\end{botherref}
\endbibitem

%%% 25
\bibitem[\protect\citeauthoryear{M.Achasov et~al.}{2023}]{STCF}
\begin{botherref}
\oauthor{\bsnm{M.Achasov}}, et al.:
Stcf conceptual design report (volume 1): Physics \& detector.
Frontiers of Physics
\textbf{19}(1)
(2023)
\doiurl{10.1007/s11467-023-1333-z}
\end{botherref}
\endbibitem

%%% 26
\bibitem[\protect\citeauthoryear{Ai et~al.}{0}]{ai}
\begin{barticle}
\bauthor{\bsnm{Ai}, \binits{X.}},
\bauthor{\bsnm{Huang}, \binits{X.}},
\bauthor{\bsnm{Li}, \binits{T.}},
\bauthor{\bsnm{Qi}, \binits{B.}},
\bauthor{\bsnm{Qin}, \binits{X.}}:
\batitle{Design and development of stcf offline software}.
\bjtitle{Modern Physics Letters A}
\bvolume{0}(\bissue{0}),
\bfpage{2440006}
(\byear{0})
\doiurl{10.1142/S0217732324400066}
\end{barticle}
\endbibitem

%%% 27
\bibitem[\protect\citeauthoryear{Huang et~al.}{2023}]{Huang_2023}
\begin{barticle}
\bauthor{\bsnm{Huang}, \binits{W.H.}},
\bauthor{\bsnm{Li}, \binits{T.}},
\bauthor{\bsnm{Li}, \binits{Q.Y.}},
\bauthor{\bsnm{Li}, \binits{H.}},
\bauthor{\bsnm{Liu}, \binits{D.}},
\bauthor{\bsnm{Huang}, \binits{X.T.}}:
\batitle{Offline software framework for the super tau charm facility}.
\bjtitle{Journal of Physics: Conference Series}
\bvolume{2438}(\bissue{1}),
\bfpage{012054}
(\byear{2023})
\doiurl{10.1088/1742-6596/2438/1/012054}
\end{barticle}
\endbibitem

\end{thebibliography}
%\input sn-article.bbl

\end{document}